\documentclass[fleqn,usenatbib]{mnras}

\usepackage[T1]{fontenc}
\usepackage{ae,aecompl}

\usepackage{graphicx}	
\usepackage{amsmath}	
\usepackage{amssymb}	
\usepackage{newtxtext,newtxmath}

\newcommand{\Msun}{\mbox{$\mathrm{M}_{\odot}$}}
\newcommand{\Rsun}{\mbox{$\mathrm{R}_{\odot}$}}

\title[Doppler tomography of the disc at HE\,1349--2305]{Velocity-imaging the rapidly precessing planetary disc around the white dwarf HE\,1349--2305 using Doppler tomography}

\author[C.\,J.\,Manser et al.]{Christopher J. Manser\,$^{1,2}$\,{\Huge \footnotemark},
Erik Dennihy\,$^{3}$,
Boris T. G\"ansicke\,$^{2,4}$,
John H. Debes\,$^{5}$, \newauthor
Nicola P. Gentile Fusillo\,$^{6}$, 
J. J. Hermes\,$^{7}$,
Mark Hollands\,$^2$,
Paula Izquierdo\,$^{8,9}$, 
B. C. Kaiser\,$^{10}$, \newauthor
T. R. Marsh\,$^2$,
Joshua S. Reding\,$^{10}$,
Pablo Rodríguez-Gil$^{8,9}$,
Dimitri Veras\,$^{2,4,11}$,
David J. Wilson\,$^{12}$\\
$^{1}$ Astrophysics Group, Department of Physics, Imperial College London, Prince Consort Rd, London, SW7 2AZ, UK \\
$^{2}$ Department of Physics, University of Warwick, Coventry CV4 7AL, UK \\
$^{3}$ Gemini Observatory\,/\,NSF’s NOIRLab, Casilla 603, La Serena, Chile \\
$^{4}$ Centre for Exoplanets and Habitability, University of Warwick, Coventry, UK \\
$^{5}$ AURA for ESA, Space Telescope Science Institute, Baltimore, MD, USA \\
$^{6}$ European Southern Observatory, Karl-Schwarzschild-Str. 2, D 85748 Garching, Germany \\
$^{7}$ Boston University, Department of Astronomy, Boston, MA 02215, USA \\
$^{8}$ Instituto de Astrof\'isica de Canarias, E-38205 La Laguna, Tenerife, Spain\\
$^{9}$ Universidad de La Laguna, Departamento de Astrof\'isica, E-38206 La Laguna, Tenerife, Spain\\
$^{10}$ University of North Carolina at Chapel Hill, Department of Physics and Astronomy, Chapel Hill, NC 27599, USA \\
$^{11}$ STFC Ernest Rutherford Fellow \\
$^{12}$ McDonald Observatory, University of Texas at Austin, Austin, TX 78712, USA \\
}

\date{Accepted XXX. Received YYY; in original form ZZZ}

\pubyear{2020}

\begin{document}
\label{firstpage}
\pagerange{\pageref{firstpage}--\pageref{lastpage}}
\maketitle

\begin{abstract}
The presence of planetary material in white dwarf atmospheres, thought to be accreted from a dusty debris disc produced via the tidal disruption of a planetesimal, is common. Approximately five\,per\,cent of these discs host a co-orbital gaseous component detectable via emission from atomic transitions\,--\,usually the 8600\,\AA\ Ca\,{\textsc{ii}} triplet. These emission profiles can be highly variable in both morphology and strength. Furthermore, the morphological variations in a few systems have been shown to be periodic, likely produced by an apsidally precessing asymmetric disc. Of the known gaseous debris discs, that around HE\,1349--2305 has the most rapidly evolving emission line morphology, and we present updated spectroscopy of the Ca\,{\textsc{ii}} triplet of this system. The additional observations show that the emission line morphologies vary periodically and consistently, and we constrain the period to two aliases of 459\,$\pm$\,3\,d and 502\,$\pm$\,3\,d. We produce images of the Ca\,{\textsc{ii}} triplet emission from the disc in velocity space using Doppler tomography\,--\,only the second such imaging of a white dwarf debris disc. We suggest that the asymmetric nature of these velocity images is generated by gas moving on eccentric orbits with radially-dependent excitation conditions via photo-ionisation from the white dwarf. We also obtained short-cadence ($\simeq$\,4\,min) spectroscopy to search for variability on the time-scale of the disc's orbital period ($\simeq$\,hours) due to the presence of a planetesimal, and rule out variability at a level of $\simeq$\,1.4\,per\,cent.

\end{abstract}


\footnotetext{E-mail: c.j.manser92@googlemail.com}



\section{Introduction}

The observational evidence of planetary systems around white dwarfs is abundant \citep{zuckerman+becklin87-1, grahametal90-1, aannestadetal93-1, vanderburgetal15-1, gaensickeetal19-1, manseretal19-1, vanderboschetal19-1, vanderburgetal20-1, guidryetal21-1, vanderboschetal21-1}. To date, the majority of detections of planetary material at white dwarfs have come from identifying pollutant metals in their atmospheres - the result of accreting planetary material (rocky planetesimals\footnote{We define `planetesimal' to refer to a planetary body between $\sim$\,1\,km to several 100\,km in size.} \citealt{zuckermanetal03-1, zuckermanetal10-1, koesteretal14-1}, icy Kuiper-belt analogues \citealt{xuetal17-1}, and even giant planet atmospheres \citealt{barstowetal14-1, gaensickeetal19-1, schreiberetal19-1}), with $\simeq$\,50\,per\,cent of white dwarfs showing evidence of hosting remains of planetary systems. The majority of detections are thought to involve planetesimals that are assumed to have been perturbed onto highly eccentric orbits ($e > 0.98$) by more massive bodies in the system \citep{debesetal02-1}, bringing them within the tidal disruption radius of the white dwarf ($\simeq$\,1\,\Rsun). The planetesimal is then ripped apart, forming a highly eccentric debris stream that is assumed to circularise within the tidal disruption radius to form a disc of dusty debris \citep{jura03-1, debesetal12-1, verasetal14-1, verasetal15-1, malamud+perets20-1, malamud+perets20-2}. These debris discs are usually detected via their infrared emission which is in excess of the white dwarf continuum \citep{zuckerman+becklin87-1, rocchettoetal15-1, farihi16-1, chenetal20-1, dennihyetal20-1, dennihyetal20-2, fusilloetal20-1, xuetal20-1}. 

Five per cent of infrared-bright debris discs around white dwarfs are observed to host an additional gaseous component in emission \citep{gaensickeetal06-3, manseretal20-1}, identified via the double-peaked Ca\,{\textsc{ii}} 8600\,\AA\ triplet emission profiles produced by a flat, photo-ionised Keplerian disc \citep{horne+marsh86-1, melisetal10-1, kinnear11, gaensickeetal19-1}. The origin of these gaseous components is uncertain, but current mechanisms include: runaway sublimation of dust at the inner edge of the debris disc due to angular momentum conservation \citep{rafikov11-2, metzgeretal12-1}, a collisional cascade of rocky bodies being ground down into dust and gas \citep{kenyon+bromley17-1, kenyon+bromley17-2}, collisions produced via a tidal stream of planetary debris impacting on a pre-existing disc \citep{jura08-1, malamudetal21-1}, and a disc-embedded planetesimal that survived the tidal disruption process, inducing the production of gas through collisions or sublimation \citep{manseretal19-1, trevascusetal21-1}. Recent observations show that variability of the infrared excess from debris discs is common \citep{xu+jura14-1, xuetal18-1, swanetal19-2, wangetal19-1}, and it has been proposed that the observed variations are due to the production and destruction of dust via planetesimal collisions which could also produce observable gaseous material \citep{Farihietal18-1, swanetal21-1}. This is further corroborated by the discovery that debris discs with a gaseous component in emission appear to show the largest amounts of infrared variability \citep{swanetal20-1}, indicating they are the most dynamically active white dwarf debris discs. Henceforth we refer to dusty debris discs that have an additional, co-orbital gaseous component in emission as `gaseous debris discs'.

The gaseous emission profiles encode velocity information within the disc which provides insight into its dynamical structure. Long-term observations of the Ca\,{\textsc{ii}} triplet profiles show similar morphological variations of the emission profiles \citep{gaensickeetal08-1, melisetal10-1, wilsonetal15-1, manseretal16-2, dennihyetal20-2, fusilloetal20-1}, which have been well-modelled for the disc around SDSS\,J122859.93+104032.9 (hereafter SDSS\,J1228+1040) using Doppler tomography \citep{marsh+horne88-1, manseretal16-1}. This modelling showed that the observed variability can be reproduced by the precession of a fixed, asymmetric intensity pattern in the disc on a period of $\simeq$\,27\,yr. This is in agreement with theoretical studies which have shown that eccentric gas discs around white dwarfs should precess (with no radial dependence) due to a combination of general relativistic precession and gas pressure forces \citep{miranda+rafikov18-1}.

HE\,1349--2305 is a white dwarf of spectral type DBAZ \citep{koesteretal05-2}, with an atmosphere dominated by helium harbouring trace amounts of hydrogen, as well as trace metals ingested from a debris disc that is detected as an infrared excess \citep{girvenetal12-1}. The disc around HE\,1349--2305 also hosts a gaseous component that appears to vary in a similar manner to SDSS\,J1228+1040, albeit with a tentative period of just 1.4\,$\pm$\,0.2\,yr (511\,$\pm$\,73\,d) \citep{melisetal12-1, dennihyetal18-1}. This period is an order-of-magnitude shorter than that of the disc around SDSS\,J1228+1040 (which has not yet been observed for a single full precession cycle), and gives us the opportunity to study the variability of these discs over multiple cycles to establish whether this period is stable. We note that the gaseous material around the white dwarf WD\,1145+017 has been modelled as an eccentric disc that precesses with a period of 4.6\,yr \citep{cauleyetal18-1, fortinetal20-1}, however this gaseous component is only detected via absorption and shows no evidence of emission.

In this manuscript we extend the $\simeq$\,0.95\,yr of spectroscopy of the gaseous debris disc around HE\,1349--2305 presented by \cite{dennihyetal18-1} by $\simeq$1.84\,yr, covering now a total of $\simeq$\,2 cycles, in addition to two archival observation obtained in 2009 and 2011. We reanalyse the variability of the Ca\,{\textsc{ii}} triplet using both a time series analysis of their velocity centroids, and Doppler tomography confirming the disc is undergoing apsidal precession. We also inspect $\simeq$\,4\,min cadence spectroscopy acquired to search for short-term variability due to the presence of a planetesimal. Finally, we discuss the results and the prospects of using Doppler maps as key probes in understanding the structure and evolution of gaseous debris discs.

\section{Observations}

The observations presented in this section were obtained from five observatories with the goal of monitoring the Ca\,{\textsc{ii}} triplet. The observations are documented in Table\,\ref{t-dates}, and details of the instruments and telescopes used are given below.

\begin{table*}
\centering
\caption{Log of HE\,1349--2305 observations. The exposure times for the UVB/VIS arms of X-shooter are given separately.  \label{t-dates}}
\begin{tabular}{llllrrr}
\hline
Date       & MJD [d] & Telescope/Instrument        & Wavelength range [\AA] & Total exposure time [s] & Resolution [\AA] \\
\hline                              
2009-05-20  & 54971.2 & Magellan/MIKE               &  3350\,--\,9500             & 1400              & 0.3\\ 
2011-05-27  & 55708.0 & VLT/X-shooter               &  3080\,--\,10\,400          & 5900\,/\,5680     & 1.0\\ 
2016-08-21  & 57621.0 & SOAR/Goodman                &  7900\,--\,9000             & 7500              & 2.2\\ 
2017-01-12  & 57765.3 & SOAR/Goodman                &  7900\,--\,9000             & 3600              & 2.2\\ 
2017-01-25  & 57778.4 & SOAR/Goodman                &  7900\,--\,9000             & 3600              & 2.2\\ 
2017-02-05  & 57789.0 & GTC/OSIRIS                  &  7350\,--\,10\,150          & 1800              & 2.5\\ 
2017-02-09  & 57793.3 & SOAR/Goodman                &  7900\,--\,9000             & 3600              & 2.2\\ 
2017-02-17  & 57801.3 & SOAR/Goodman                &  7900\,--\,9000             & 8400              & 2.2\\ 
2017-03-10  & 57822.2 & SOAR/Goodman                &  7900\,--\,9000             & 5400              & 2.2\\ 
2017-03-14  & 57826.2 & SOAR/Goodman                &  7900\,--\,9000             & 3600              & 2.2\\ 
2017-04-11a & 57854.2 & SOAR/Goodman                &  7900\,--\,9000             & 5400              & 2.2\\
2017-04-11b & 57854.3 & SOAR/Goodman                &  7900\,--\,9000             & 7200              & 2.2\\ 
2017-04-23  & 57866.2 & SOAR/Goodman                &  7900\,--\,9000             & 3600              & 2.2\\ 
2017-05-30  & 57903.0 & SOAR/Goodman                &  7900\,--\,9000             & 7200              & 2.2\\ 
2017-06-09  & 57913.1 & SOAR/Goodman                &  7900\,--\,9000             & 7800              & 2.2\\ 
2017-07-25  & 57959.0 & SOAR/Goodman                &  7900\,--\,9000             & 5400              & 2.2\\ 
2017-08-02  & 57968.0 & SOAR/Goodman                &  7900\,--\,9000             & 6600              & 2.2\\ 
2018-02-04  & 58153.3 & SOAR/Goodman                &  7900\,--\,9000             & 5400              & 2.2\\ 
2018-02-10  & 58159.3 & SOAR/Goodman                &  7900\,--\,9000             & 4200              & 2.2\\ 
2018-03-27  & 58204.3 & SOAR/Goodman                &  7900\,--\,9000             & 3600              & 2.2\\ 
2018-04-27  & 58235.3 & SOAR/Goodman                &  7900\,--\,9000             & 5400              & 2.2\\ 
2018-05-15  & 58253.2 & VLT/X-shooter               &  3080\,--\,10\,400          & 2950\,/\,2840     & 1.0\\ 
2018-06-02  & 58271.1 & SOAR/Goodman                &  7900\,--\,9000             & 9000              & 2.2\\ 
2019-03-09  & 58551.3 & Gemini South/GMOS-S         &  8260\,--\,8790             & 9840              & 3.0\\ 
2019-03-16  & 58558.2 & Gemini South/GMOS-S         &  8260\,--\,8790             & 10\,800           & 3.0\\ 
2019-03-24  & 58566.2 & SOAR/Goodman                &  7900\,--\,9000             & 7200              & 2.2\\ 
2019-04-23  & 58596.1 & SOAR/Goodman                &  7900\,--\,9000             & 6000              & 2.2\\ 
2019-05-15  & 58618.2 & SOAR/Goodman                &  7900\,--\,9000             & 7200              & 2.2\\ 
2019-06-06  & 58640.1 & VLT/X-shooter               &  3080\,--\,10\,400          & 2950\,/\,2840     & 1.0\\ 
\hline
\end{tabular}
\end{table*}

\textit{Southern Astrophysical Research (SOAR) Telescope/Goodman High Throughput Spectrograph (Goodman)}: Goodman is a highly configurable low resolution spectrograph mounted on the SOAR 4m telescope \citep{clemensetal04-1}. The data were collected using the 1200l-R grating and were reduced using a custom set of \textsc{python}-based tools described in \citep{dennihyetal18-1}, utilising an optimal extraction routine based on the methods described in \cite{marsh89-1}.

\begin{figure*}
\centerline{\includegraphics[width=2\columnwidth]{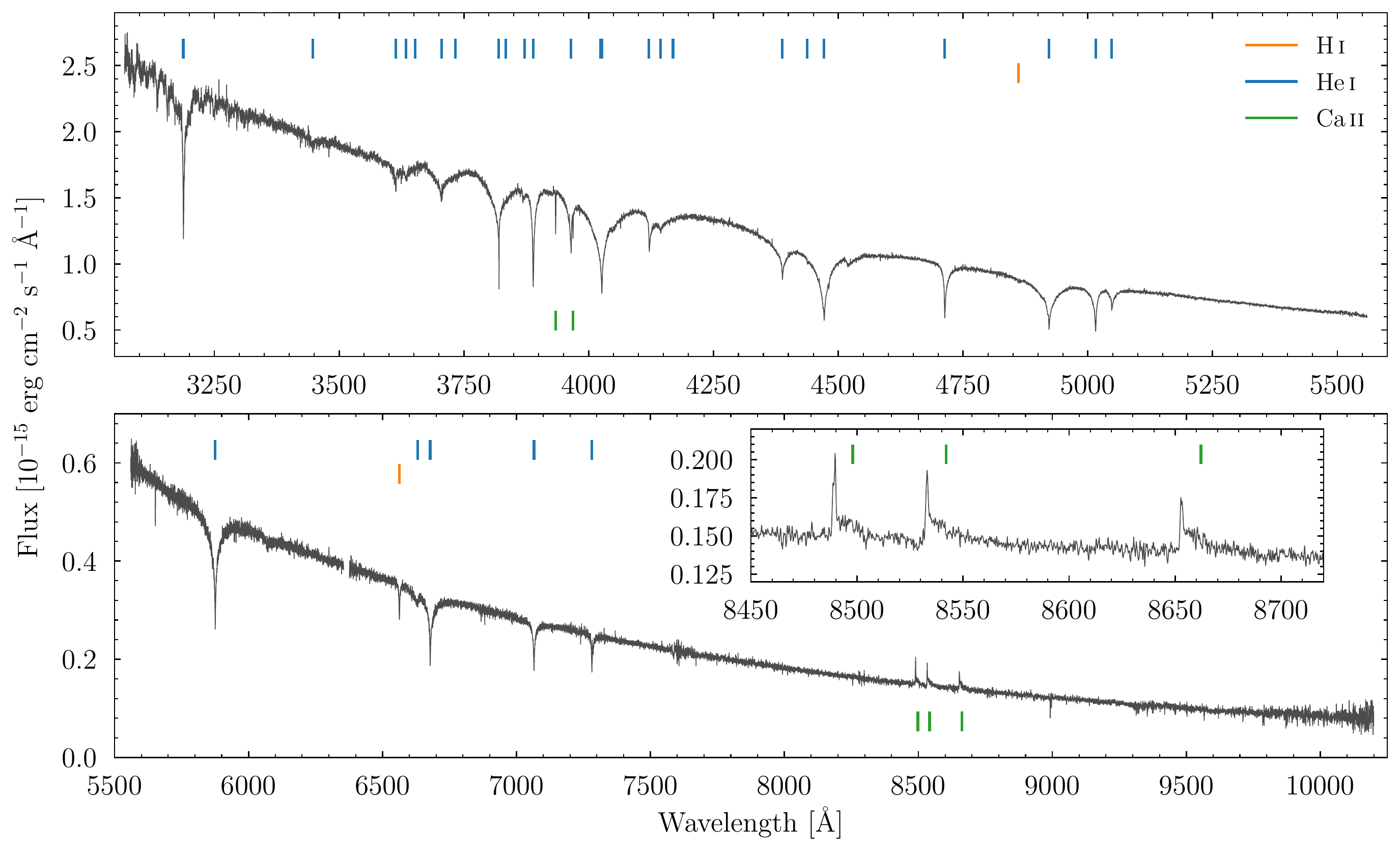}}
\caption{\label{f-vis_spec} The 2019 June X-shooter spectrum of HE\,1349--2305. The inset shows a zoom-in of the Ca\,{\textsc{ii}} triplet region, revealing a blue-dominated circumstellar emission feature, with little red-shifted material in emission implying a strongly asymmetric brightness distribution in the disc. The coloured tabs denote the rest wavelengths of strong H, He, and Ca features; only the lowest two energy transitions of the Balmer series are detected.}
\end{figure*}

\textit{Very Large Telescope (VLT)/X-shooter:} X-shooter is an intermediate resolution echelle spectrograph on the ESO VLT \citep{vernetetal11-1}. The data (program IDs: 087.D-0858(A), 5100.C-0407(C), and 5100.C-0407(I)) were reduced within the \textsc{reflex}\,\footnote{Documentation and software for \textsc{reflex} can be obtained from http://www.eso.org/sci/software/reflex/} reduction work flow using the standard settings and optimising the slit integration limits \citep{freudlingetal13-1}, and a telluric correction was performed using \textsc{molecfit} \citep{smetteetal15-1, kauschetal15-1}. The 2019 June spectrum is shown in Fig.\,\ref{f-vis_spec}, where the NIR arm of X-shooter was not used due to low counts in the infrared.

\textit{Magellan/Magellan Inamori Kyocera Echelle (MIKE):} The MIKE spectrograph on the Magellan telescope is a double-echelle spectrograph capable of high resolution over a wide wavelength range \citep{bernsteinetal03-1}. The data were reduced and extracted using the Carnegie \textsc{python} tools MIKE data reduction pipeline, with methodology described in \cite{kelson03-1}.

\textit{Gran Telescopio Canarias (GTC)/Optical System for Imaging and low-Intermediate-Resolution Integrated Spectroscopy (OSIRIS):} HE\,1349--2305 was observed at the 10.4\,m GTC on 2017 Feburary 5 using the OSIRIS spectrograph with the volume-phased holographic R2500I grating (program ID: GTC1-16ITP). The observations were reduced using standard techniques under the \textsc{starlink}\footnote{The \textsc{starlink} and \textsc{pamela} software are available at \url{http://starlink.eao.hawaii.edu/starlink}} software package. The science frames were bias-subtracted and flat-fielded, and sky-subtraction and extraction of the 1-D spectra were performed using the \textsc{pamela}\footnotemark[2] software package, where the optimal-extraction algorithm was used to maximise the spectral signal-to-noise ratio. The \textsc{molly}\footnote{\textsc{molly} software is available at \url{http://deneb.astro.warwick.ac.uk/phsaap/software/}} package was used for wavelength calibration of the extracted 1-D data.

\textit{Gemini South/Gemini Multi-Object Spectrograph (GMOS-S)}: GMOS-S on Gemini South is a highly configurable instrument capable of multi-object, longslit and integral-field unit spectroscopy \citep{hooketal04-1,gimenoetal16-1}. The data were collected using the R-831 grating (program ID: S-2019A-Q-231). The data were reduced using a combination of the Gemini \textsc{iraf} package and an optimal extraction routine based on the methods described in \cite{marsh89-1}. We obtained a total of 86, 4\,min exposures (totalling 20640\,s) on 2019 March 9 (41 exposures) and 16 (45 exposures) to search for any short-term variations in the emission strength and morphology of the Ca\,{\textsc{ii}} triplet on the time-scale of the orbital period in the disc ($\simeq$\,hours).

\section{Gaseous emission profiles}

The characteristic emission profile from a circular gaseous disc of gas on Keplerian orbits with a radially symmetric intensity distribution is a symmetric, double-peaked emission profile (see fig.1 of \citealt{horne+marsh86-1}). These profiles can reveal structural properties of the disc such as the inner and outer radii. The inner radius of such a disc can be determined by the velocity at which the emission drops to the continuum level of the spectrum, and the absolute velocity between the blue- and red-shifted sides of the profile is known as the full width zero intensity (FWZI). A FWZI not centered at 0\,km\,s$^{-1}$ is indicative of an asymmetry in the disc, and may be due to a non-symmetric emitting area in a circular disc (e.g. spiral shocks, \citealt{steeghs+stehle99-1}), or a non-circular disc geometry (e.g. eccentric discs, \citealt{fortinetal20-1}). The outer radius of a circular disc in emission can be estimated from the peak separation of the double-peaked profile, but is also affected by disc asymmetries which can lead to differences in the positions and strengths of the two peaks. This type of asymmetry is clearly evident in the emission profiles of HE\,1349--2350 (Fig.\,\ref{f-vis_spec}), where only one emission peak is evident. The radii estimates, $R_\textrm{obs}$ obtained through these methods are affected by the inclination, $i$ of the disc, and to determine the true radii, $R$, one must use $R_\textrm{obs} = R \sin^2 i$. We use this insight to discuss the line profiles observed at HE\,1349--2305.

\begin{figure*}
\centerline{\includegraphics[width=2\columnwidth]{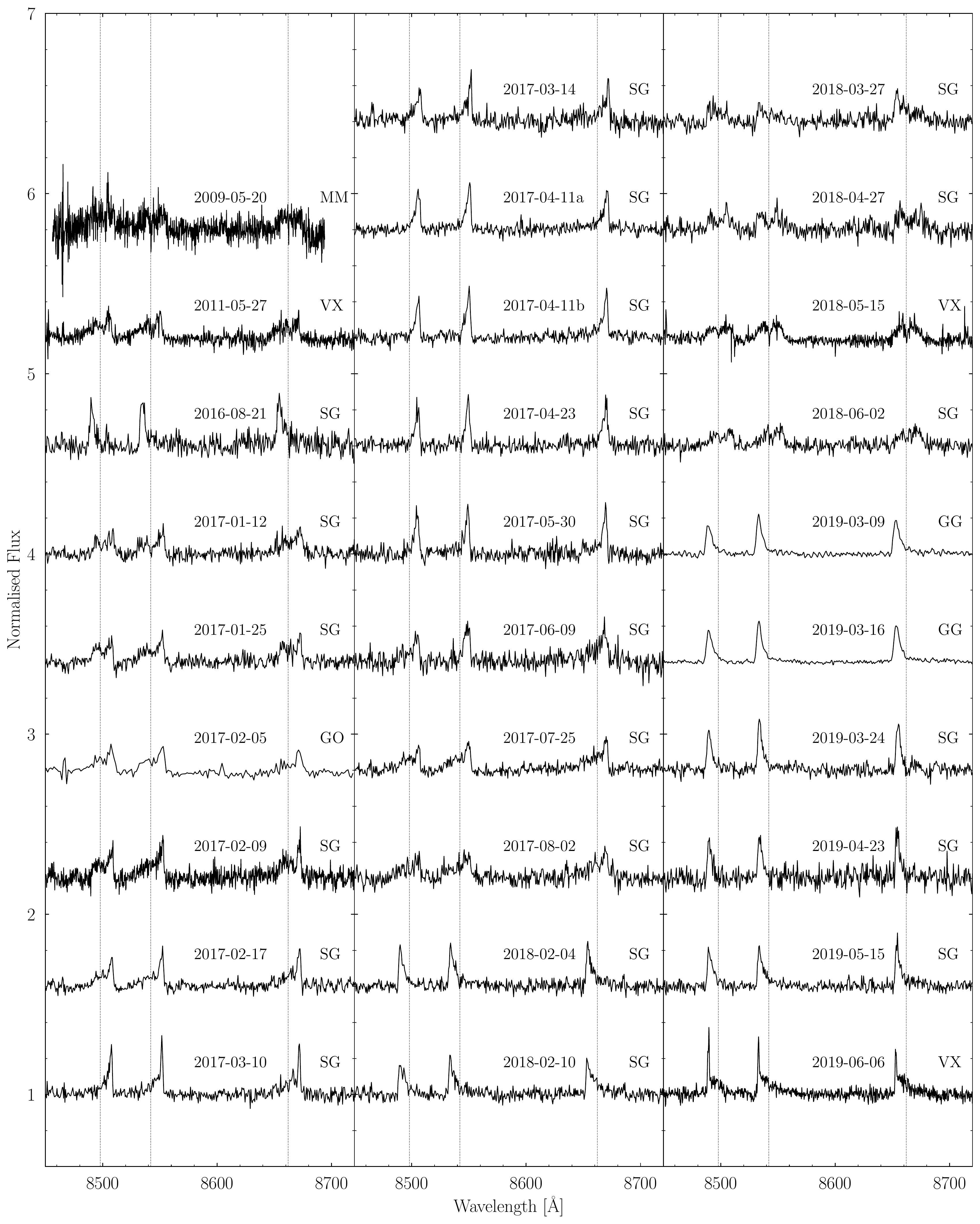}}
\caption{\label{f-caii_profiles} Continuum-normalised spectra of the Ca\,{\textsc{ii}} triplet emission from the gaseous component of the debris disc around HE\,1349--2305. The epochs of the 29 spectra are given along with a two letter code using the first initial of the telescope and instrument used for the observation, see Table\,\ref{t-dates}. The spectra are shifted in the $y$-axis by steps of 0.6 from a value of one for clarity, and vertical dashed lines correspond to the rest wavelengths of the triplet components.}
\end{figure*} 

The Ca\,{\textsc{ii}} triplet emission from the disc around HE\,1349--2305 is undergoing morphological variability as shown by the emission profiles in Fig.\,\ref{f-caii_profiles}, where the spectra are continuum-normalised in the wavelength range 8450\,--\,8720\,\AA. We have extended the intense monitoring of the Ca\,{\textsc{ii}} triplet of \protect\cite{dennihyetal18-1} from 2018 February to 2019 June with twelve additional spectra, which confirm the continued variation of the emission profiles (Fig.\,\ref{f-caii_profiles}). Below we analyse the morphological variations of the Ca\,{\textsc{ii}} triplet using two methods: (i) Velocity centroid fitting, and (ii) Doppler tomography.

\subsection{Velocity centroid fitting of the Ca\,{\textsc{ii}} triplet}

\begin{table}
\centering
\caption{Log of velocity centroids given for each component of the Ca\,{\textsc{ii}}. \label{t-velcen}}
\begin{tabular}{llrrr}
\hline
           &         & Ca\,{\textsc{ii}} 8498\,\AA\     & Ca\,{\textsc{ii}} 8542\,\AA      & Ca\,{\textsc{ii}} 8662\,\AA      \\
Date       & MJD [d] &                    \multicolumn{3}{|c|}{Velocity centroid [km\,s$^{-1}$]}                              \\
\hline                 
2011-05-27  & 55708.0 & 7     $\pm$ 24 ~  & --8   $\pm$ 19 ~  & --21  $\pm$ 15 ~ \\
2016-08-21  & 57621.0 & --218 $\pm$ 34 ~  & --215 $\pm$ 29 ~  & --215 $\pm$ 30 ~ \\
2017-01-12  & 57765.3 & 156   $\pm$ 52 ~  & 122   $\pm$ 81 ~  & 160   $\pm$ 62 ~ \\
2017-01-25  & 57778.4 & 95    $\pm$ 77 ~  & 15    $\pm$ 55 ~  & 102   $\pm$ 50 ~ \\
2017-02-05  & 57789.0 & 158   $\pm$ 27 ~  & 193   $\pm$ 45 ~  & 207   $\pm$ 49 ~ \\
2017-02-09  & 57793.3 & 107   $\pm$ 46 ~  & 92    $\pm$ 45 ~  & 167   $\pm$ 77 ~ \\
2017-02-17  & 57801.3 & 144   $\pm$ 33 ~  & 251   $\pm$ 62 ~  & 180   $\pm$ 72 ~ \\
2017-03-10  & 57822.2 & 202   $\pm$ 56 ~  & 257   $\pm$ 19 ~  & 253   $\pm$ 28 ~ \\
2017-03-14  & 57826.2 & 256   $\pm$ 41 ~  & 232   $\pm$ 44 ~  & 258   $\pm$ 65 ~ \\
2017-04-11a & 57854.2 & 229   $\pm$ 21 ~  & 246   $\pm$ 19 ~  & 247   $\pm$ 25 ~ \\
2017-04-11b & 57854.3 & 204   $\pm$ 31 ~  & 247   $\pm$ 82 ~  & 235   $\pm$ 25 ~ \\
2017-04-23  & 57866.2 & 223   $\pm$ 19 ~  & 207   $\pm$ 28 ~  & 227   $\pm$ 23 ~ \\
2017-05-30  & 57903.0 & 197   $\pm$ 17 ~  & 207   $\pm$ 21 ~  & 195   $\pm$ 24 ~ \\
2017-06-09  & 57913.1 & 208   $\pm$ 31 ~  & 86    $\pm$ 62 ~  & 161   $\pm$ 47 ~ \\
2017-07-25  & 57959.0 & 64    $\pm$ 42 ~  & 42    $\pm$ 55 ~  & 64    $\pm$ 41 ~ \\
2017-08-02  & 57968.0 & 143   $\pm$ 65 ~  & 80    $\pm$ 56 ~  & 92    $\pm$ 139  \\
2018-02-04  & 58153.3 & --216 $\pm$ 30 ~  & --208 $\pm$ 31 ~  & --204 $\pm$ 38 ~ \\
2018-02-10  & 58159.3 & --180 $\pm$ 44 ~  & --204 $\pm$ 43 ~  & --184 $\pm$ 24 ~ \\
2018-03-27  & 58204.3 & --122 $\pm$ 72 ~  & --60  $\pm$ 67 ~  & --66  $\pm$ 98 ~ \\
2018-04-27  & 58235.3 & 69    $\pm$ 142   & 71    $\pm$ 125   & 133   $\pm$ 72 ~ \\
2018-05-15  & 58253.2 & 82    $\pm$ 34 ~  & 96    $\pm$ 35 ~  & 86    $\pm$ 26 ~ \\
2018-06-02  & 58271.1 & 139   $\pm$ 115   & 136   $\pm$ 60 ~  & 136   $\pm$ 60 ~ \\
2019-03-09  & 58551.3 & --269 $\pm$ 18 ~  & --274 $\pm$ 9 ~ ~ & --256 $\pm$ 20 ~ \\
2019-03-16  & 58558.2 & --272 $\pm$ 9 ~ ~ & --251 $\pm$ 9 ~ ~ & --261 $\pm$ 10 ~ \\
2019-03-24  & 58566.2 & --271 $\pm$ 23 ~  & --212 $\pm$ 23 ~  & --231 $\pm$ 39 ~ \\
2019-04-23  & 58596.1 & --244 $\pm$ 30 ~  & --263 $\pm$ 57 ~  & --239 $\pm$ 45 ~ \\
2019-05-15  & 58618.2 & --216 $\pm$ 32 ~  & --219 $\pm$ 28 ~  & --233 $\pm$ 20 ~ \\
2019-06-06  & 58640.1 & --164 $\pm$ 23 ~  & --274 $\pm$ 27 ~  & --151 $\pm$ 26 ~ \\
\hline
\end{tabular}
\end{table}

\begin{figure*}
\centerline{
\includegraphics[width=1.1\columnwidth]{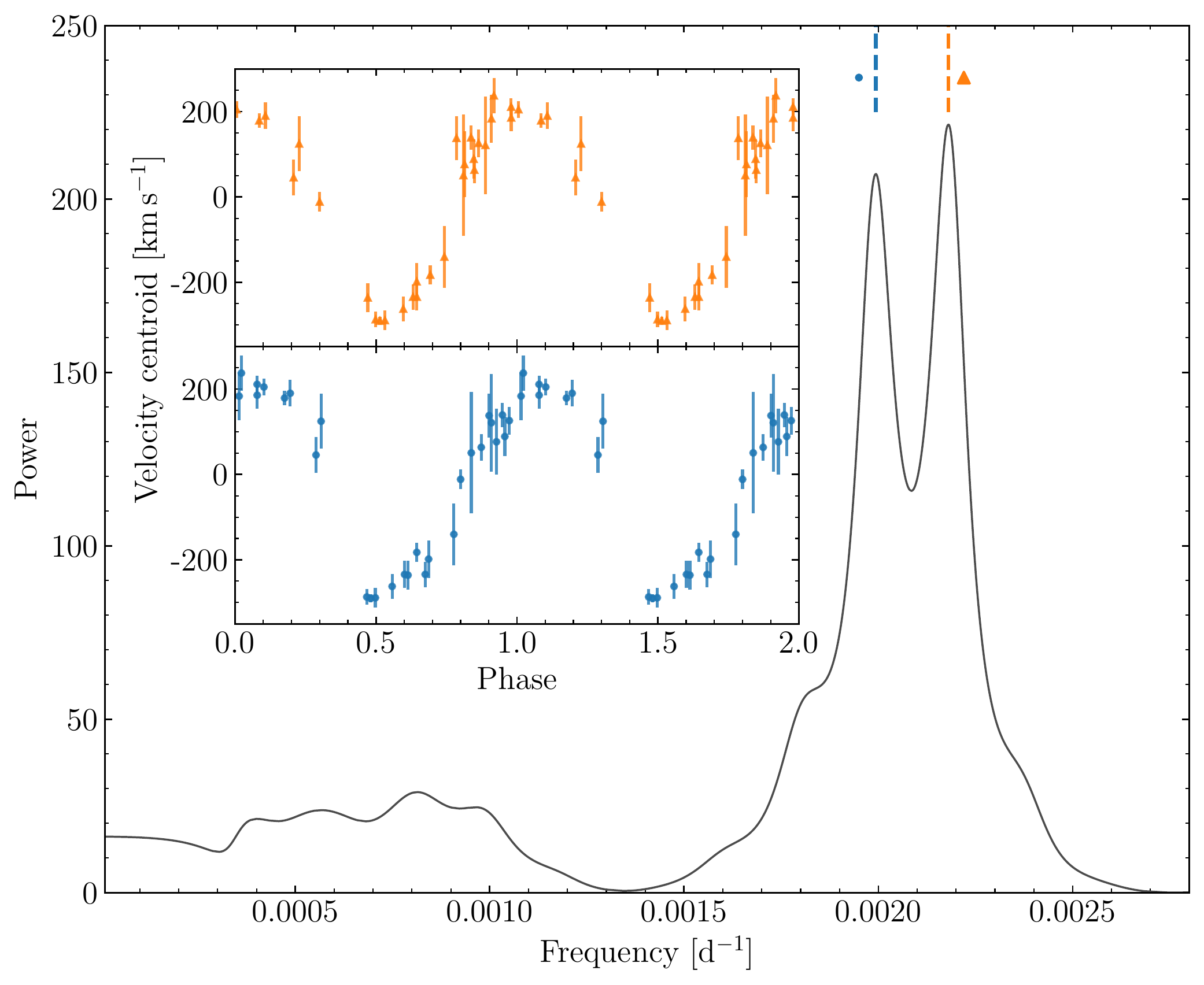}
\includegraphics[width=0.9\columnwidth]{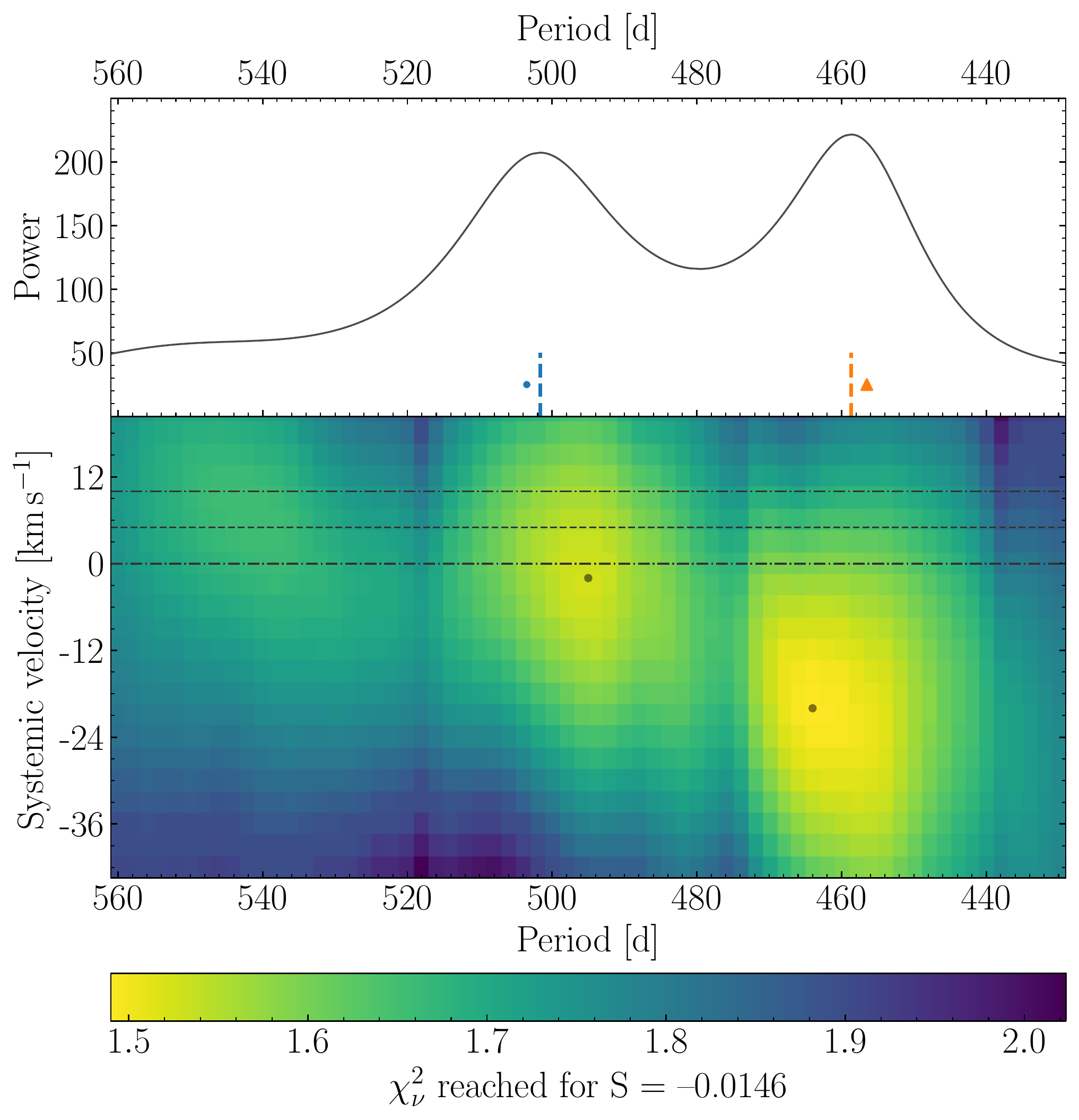}
}
\caption{\label{f-period_plots} Left: The periodogram calculated using the ANOVA method (see text) of the triplet-averaged velocity centroids of the profiles in Fig.\,\ref{f-caii_profiles} displays two distinct peaks at periods 459\,$\pm$\,3\,d (orange triangle) and 502\,$\pm$\,3\,d (blue circle). The inset shows the velocity centroids phase-folded onto the two periods using the same colours and symbols. 
Right: The top panel reproduces a section of the periodogram presented in the left panel in period space. The bottom panel shows the $\chi^2_{\nu}$, for Doppler tomograms reached at an entropy, $S$\,=\,$-0.0146$, for a range of precession periods and systemic velocities. Lower values signify better fits, with two local minima denoted by black dots. The dashed line shows the systemic velocity determined by \protect\cite{melisetal12-1}, and the dot-dashed lines depict a lower limit on the $\pm$\,1\,$\sigma$ region.}
\end{figure*}

The velocity centroid method relies on measuring the flux-weighted average velocity of the emission profiles as they oscillate between blue- and red-dominated phases\footnote{This method was also used in \cite{cauleyetal18-1} to estimate the period of precession of the gaseous material in absorption around WD\,1145+017.}. We follow the same methods as outlined in \cite{dennihyetal18-1} to measure the velocity centroid for each emission profile at every epoch\footnote{The systematic differences in the velocity centroid zero-points due to the wavelength calibration of the different spectroscopic datasets are $\lesssim$\,5\,km\,s$^{-1}$. We take this to be negligible and exclude its effects on the following analysis.} (Table\,\ref{t-velcen}), excluding the 2009 May MIKE spectrum from the analysis as its average signal-to-noise ratio of $\simeq$\,13 is not sufficient for velocity centroiding. The velocity centroids for the Ca\,{\textsc{ii}} triplet profiles were then averaged at each epoch and the resulting time-series velocity centroids were analysed using the analysis of variance (ANOVA) method, which is particularly powerful in the case of unevenly sampled data \citep[see][for details]{schwarzenberg-czerny96-1}. The resulting periodogram is shown in Fig.\,\ref{f-period_plots}, which reveals two clear peaks in frequency at (2.18\,$\pm$\,0.01)\,$\times$10$^{-3}$\,d$^{-1}$ and (1.99\,$\pm$\,0.01)\,$\times$10$^{-3}$\,d$^{-1}$ corresponding to periods of 459\,$\pm$\,3\,d (1.257\,$\pm$\,0.008\,yr) and 502\,$\pm$\,3\,d (1.374\,$\pm$\,0.009\,yr) respectively. Due to the $\simeq$\,0.4\,--\,1\,yr gaps present in our data collection, we cannot distinguish between these two period-aliases. We phase-fold the epoch-averaged velocity centroid data on these two periods (Fig.\,\ref{f-period_plots}), to illustrate the periodic nature of the variations. Both of the periods are shorter than, but consistent with, the previously determined period of 511\,$\pm$\,73\,d. 

\subsection{Doppler tomography of the Ca\,{\textsc{ii}} triplet}

Doppler tomography uses the one-dimensional velocity information encapsulated in the emission profiles to produce a two-dimensional image of the disc in velocity space -- known as a Doppler tomogram or Doppler map. This technique relies on several assumptions: (1) That all points on the disc are equally visible at all times. (2) The flux from any point fixed in the rotating frame is constant in time. (3) All motion is parallel to the orbital plane. (4) The intrinsic width of the profile from any point is negligible \citep{marsh01-2}. As these tomograms are usually made for a specific group of atomic transitions, one can use these to map the location of multiple emitting species \citep{undasanzanaetal06-1}. To determine whether the disc around HE\,1349--2305 is also apsidally precessing with a fixed period, we compute Doppler maps for the system's gaseous emission. This also allows for the period of variability of the system to be independently calculated for comparison with the velocity centroid analysis described above.

To generate a Doppler map with the 29 normalised spectra shown in Fig.\,\ref{f-caii_profiles} we fix two free parameters: the period, $P$, and systemic velocity, $\gamma$. We use the maximum entropy inversion method described in \cite{marsh+horne88-1} whereby for a given target reduced chi-squared $\chi^2_{\nu}$, the Doppler map with the highest entropy $S$ (or least information) is selected. As the target $\chi^2_{\nu}$ is reduced, the maximum value of $S$ will also decrease from a value of zero as a more complex structure in the Doppler map is required. As such, for a given value of $P$ and $\gamma$, a target $S$ can be selected and the corresponding map with the minimum $\chi^2_{\nu}$ can be determined. The exact value of $S$ is unimportant, and a choice of $S$\,=\,--0.0146 (a dimensionless quantity) was selected based on preliminary Doppler map fits. We computed Doppler maps for a grid of $430$\,$\textrm{d}\leq P \leq 560$\,d in steps of 2\,d and $-42 \textrm{km\,s}^{-1} \leq \gamma \leq 20$km\,s$^{-1}$ in steps of 3\,km\,s$^{-1}$, with the results shown in Fig.\,\ref{f-period_plots}.

\begin{figure*}
\centerline{\includegraphics[width=2\columnwidth]{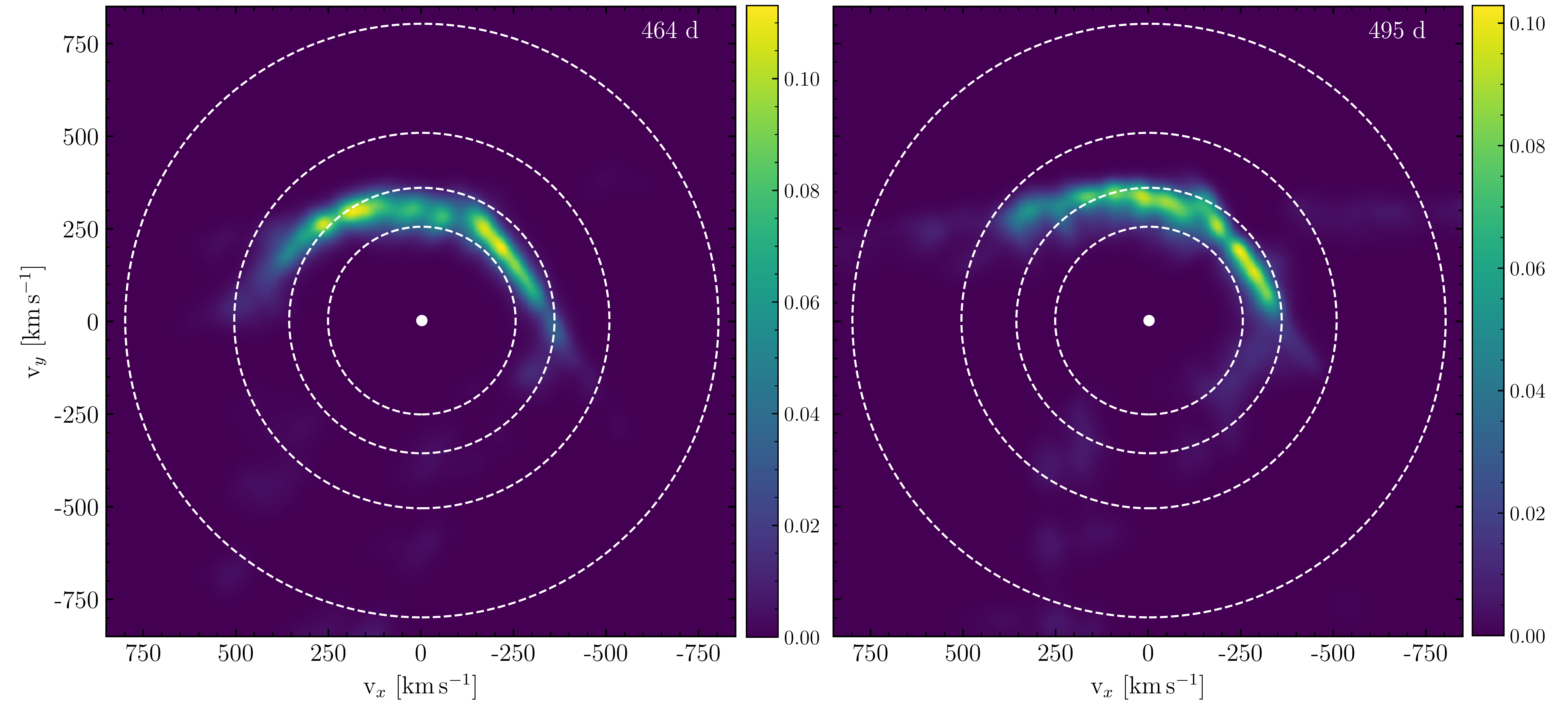}}
\caption{\label{f-doppler_464} Doppler maps of the disc around HE\,1349--2305 assuming a period of 464\,d and a systemic velocity of $-20$\,km\,s$^{-1}$ in the left panel, and a period of 495\,d and a systemic velocity of $-2$\,km\,s$^{-1}$ in the right panel. Circular orbits (white dashed lines) assuming the disc has an inclination, $i$\,=\,90\,$^{\textrm{o}}$, and the white dwarf has a mass $M_{\textrm{WD}}$\,=\,0.673\,\Msun\ \citep{koesteretal05-2, vossetal07-1}, are shown with radii 0.2, 0.5, 1.0, and 2.0 \Rsun\ in order from the largest circle to the smallest. The colour bar shows the brightness of the disc in arbitrary units.}
\end{figure*}

\begin{figure*}
\centerline{\includegraphics[width=2\columnwidth]{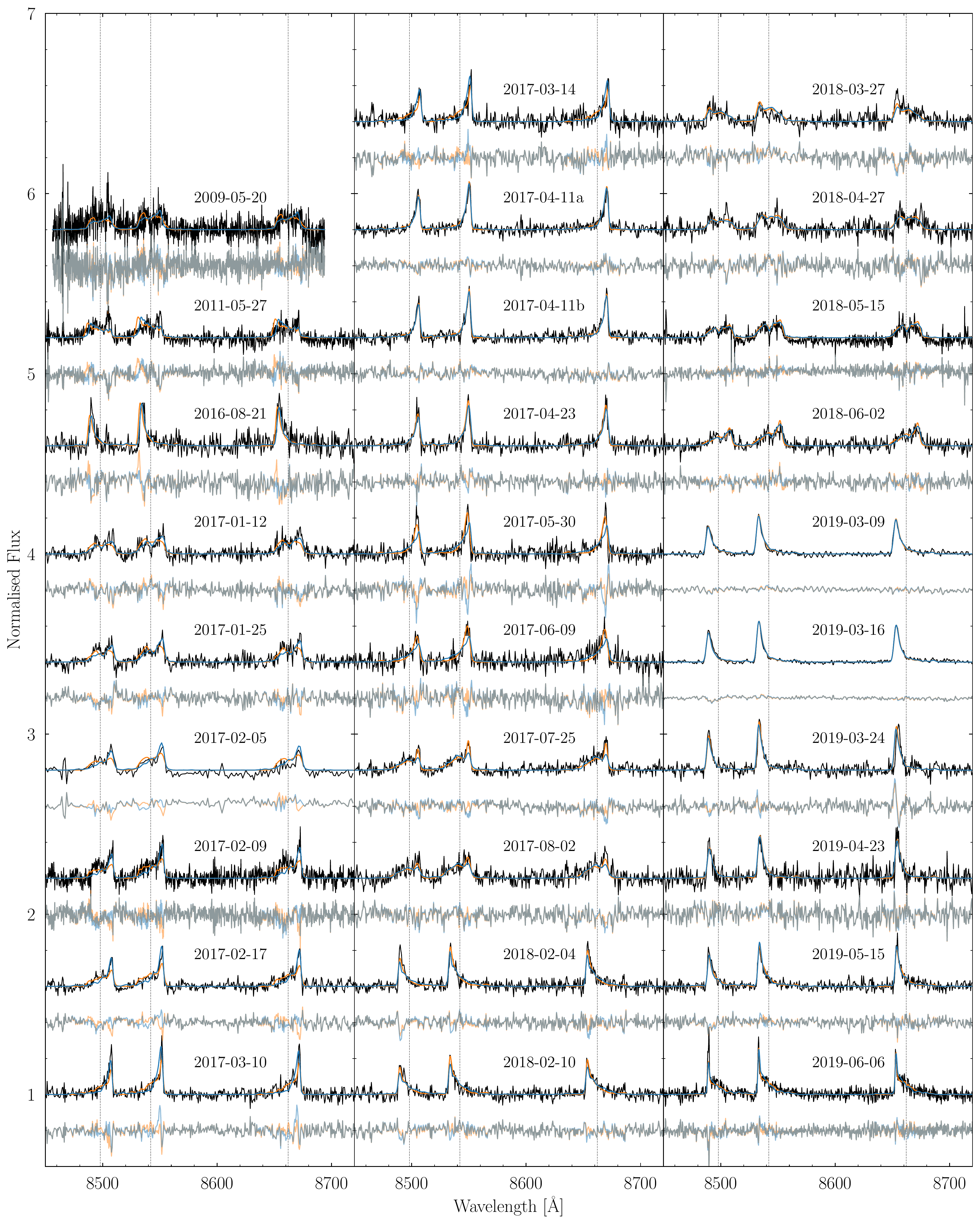}}
\caption{\label{f-profile_fits} Comparison of the observed spectra of HE\,1349--2305 and those generated from the fitted Doppler maps. The Ca\,{\textsc{ii}} triplet profiles (black) are shown with the profile fits generated from the 464\,d (orange) and 495\,d (blue) period Doppler maps presented in Fig.\,\ref{f-doppler_464}. The residuals for both fits are shown directly below each profile in orange and blue for the 464\,d and 495\,d periods respectively, with overlapping residuals presented in grey.}
\end{figure*}

We identify two minima in our $P$-$\gamma$ grid with periods of 464\,d and 494\,d, with corresponding systemic velocities of $-21$\,km\,s$^{-1}$ and $-3$\,km\,s$^{-1}$, and we perform a finer grid search around these minima moving in steps of 1\,d in $P$ and 1\,km\,s$^{-1}$ in $\gamma$. The Doppler maps produced at these minima have periods of 464\,d (1.27\,yr) and 495\,d (1.36\,yr), with corresponding systemic velocities of $-20$\,km\,s$^{-1}$ and $-2$\,km\,s$^{-1}$ and $\chi^2_{\nu}$ of 1.492 and 1.528 respectively. The precession periods we determine are in agreement with those determined by the ANOVA analysis of the velocity centroids. We show the Doppler maps corresponding to periods of 464\,d and 495\,d in Fig.\,\ref{f-doppler_464}, and present the resulting fits from these maps to the Ca\,{\textsc{ii}} triplet emission profiles in Fig.\,\ref{f-profile_fits}. The maps look similar, with an extremely asymmetric intensity distribution with only one side of the disc showing emission. We note that while these intensity patterns are fixed in a frame of constant apsidal precession, the orbital period of the material in the disc ($\simeq$\,hours) is orders of magnitude faster. The observed asymmetry is not unexpected from the emission profiles, where the most extreme epochs show the complete disappearance of one peak. For example, in 2017 April (2019 March) the blue(red)-shifted peak is not present. On both maps there are high velocity, low intensity regions of emission that are most likely artefacts arising from the assumptions of Doppler tomography being broken (see section 4.3 and fig.\,6 of \citealt{marsh+horne88-1} for examples), as they differ significantly between the two maps. Some of these can be seen at $v_{\textrm{y}}$\,$<$\,$-$250\,km\,s$^{-1}$ in both maps, and also along $v_{\textrm{y}}$\,$\simeq$\,250\,km\,s$^{-1}$ in the 495\,d period map.

Inspecting the profile fits (Fig.\,\ref{f-profile_fits}) for both periods show that an apsidally precessing disc of fixed-intensity gives an excellent qualitative explanation for the variability we see, however there are epochs for both periods where the fits deviate from the line profiles. The majority of these deviations appear to be due to slight morphological differences in the fits resulting in shallow residual profiles (e.g. 2017 February 9), or small differences in the positions of the emission peak between the data and the fit, leading to large residual spikes (e.g. 2017 May 30 and 2019 March 24). The fits of the profiles produced by the 464\,d period are worst in the earliest epochs, from 2009 May to 2017 March, after which the fits improve and are consistent over $\simeq$\,1.75 precession cycles. Comparatively the worst-fitted profiles for the 495\,d map occur at 2009 May, 2011 May, and between 2017 March and 2017 July. 

Both maps poorly fit the first two epochs in 2009 May and 2011 May - where in the latter the observed and fitted profiles appear in anti-phase for both maps. These spectra were not obtained as part of our dedicated monitoring, and are $\simeq$\,6\,--\,7 full precession cycles away from the most recent observation for both periods, and may suggest that, over multiple precession cycles, the assumptions of Doppler tomography requiring either the intensity pattern or the precession period of the disc to be constant breaks down. Similar morphological deviations of the fits are seen for SDSS\,J1228+1040 in fig.\,7 of \cite{manseretal16-1} for the first five years of data. Short-term variations (compared to the precession period of $\simeq$\,27\,yr) were invoked to account for the deviations, which may be due to the recently observed presence of a disc-embedded planetesimal \citep{manseretal19-1}. 

Due to the aliasing issues revealed by the velocity centroid analysis, the maps are very similar in both brightness and morphology, and it is not possible to reject a map based on their $\chi^2_{\nu}$ values alone. The systemic velocity has previously been estimated at $\gamma \simeq 5$\,km\,s$^{-1}$ by \cite{melisetal12-1}. The authors used data obtained by the Ultraviolet and Visual Echelle Spectrograph (UVES) on the VLT \citep{vossetal07-1} to determine the velocity shift of photospheric absorption features in the white dwarf spectrum, found to be 40\,$\pm$\,5\,km\,s$^{-1}$, and subtracted from this an estimation of the gravitational red-shift due to the white dwarf of 35\,km\,s$^{-1}$. We show this estimation of $\gamma$ on Fig.\,\ref{f-period_plots} with an error of $\pm$\,5\,km\,s$^{-1}$ from the velocity shift determination. This error, however, does not take into account the uncertainty on the gravitational red-shift estimate used. As such, we cannot currently distinguish between the two consistent periods determined using either velocity centroids or Doppler tomography.

\subsection{Additional candidate emission profiles}

\begin{figure}
\centerline{\includegraphics[width=1\columnwidth]{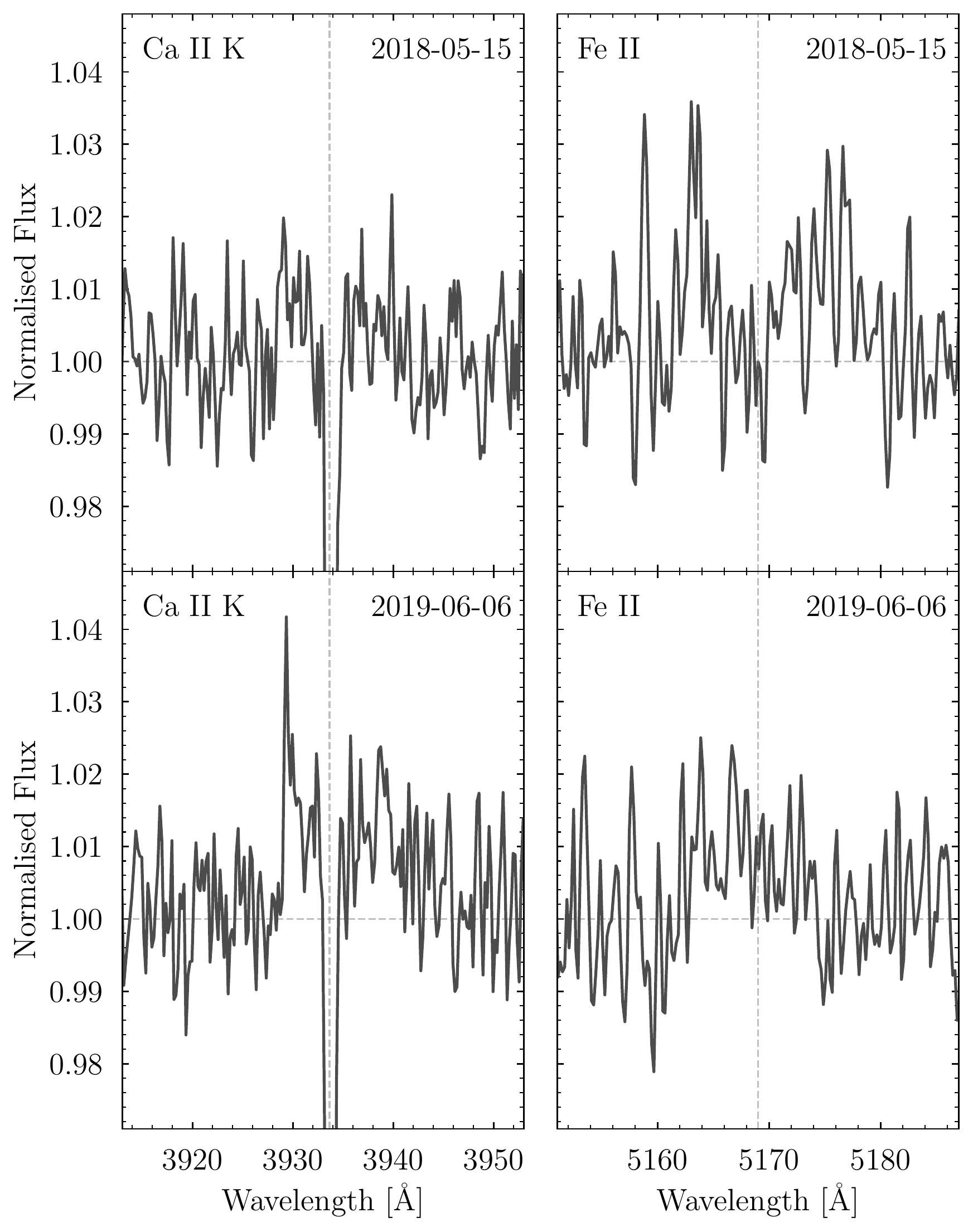}}
\caption{\label{f-CaK+Fe_spec} The 3934\,\AA\ Ca\,{\textsc{ii}}\,K and 5169\,\AA\ Fe\,{\textsc{ii}} regions of the 2018 and 2019 continuum-normalised X-shooter spectra. The absorption feature due to Ca\,{\textsc{ii}}\,K in the white dwarf photosphere is clearly present. Vertical dashed lines corresponding to zero-velocity air wavelengths of the Ca\,{\textsc{ii}}\,K and Fe\,{\textsc{ii}} transitions are presented for clarity.}
\end{figure}

\begin{figure}
\centerline{\includegraphics[width=1\columnwidth]{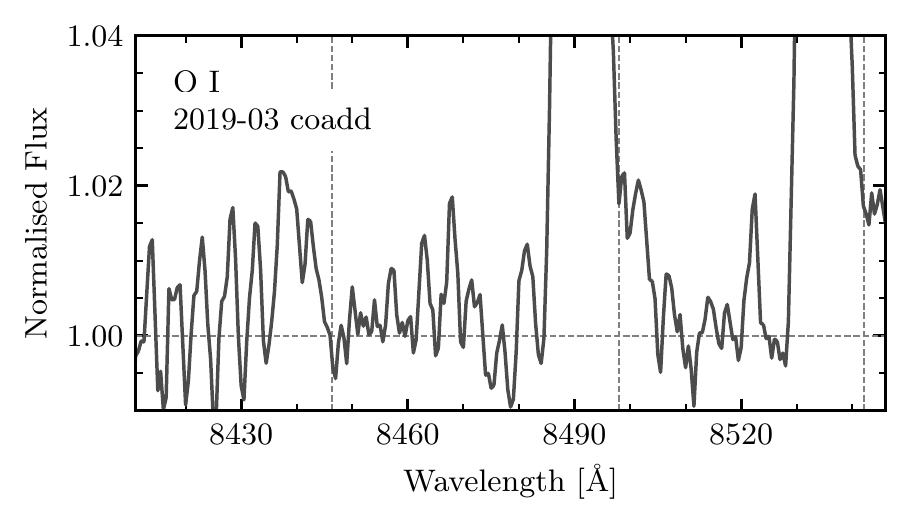}}
\caption{\label{f-OI} The 8446\,\AA\ O\,{\textsc{i}} region bluewards of the Ca\,{\textsc{ii}} triplet reveals potential emission in the coadded continuum-normalised GMOS spectra. Vertical dashed lines corresponding to zero-velocity air wavelengths of the O\,{\textsc{i}} and Ca\,{\textsc{ii}} triplet transitions are presented for clarity.}
\end{figure}

While the Ca\,{\textsc{ii}} triplet emission features were discovered and analysed in other works, we have identified subtle, candidate emission features at the 3934\,\AA\ Ca\,{\textsc{ii}}\,K, 5169\,\AA\ Fe\,{\textsc{ii}} (Fig.\,\ref{f-CaK+Fe_spec}), and 8446\,\AA\ O\,{\textsc{i}} (Fig.\,\ref{f-OI}) lines. While the 3934\,\AA\ Ca\,{\textsc{ii}}\,K feature is significantly weaker, the emission morphology appears to evolve between the two epochs from a symmetric profile to blue-shift dominated one, matching the morphological evolution of the Ca\,{\textsc{ii}} triplet emission. Furthermore, the velocity of the blue-shifted peak of the 2019 June Ca\,{\textsc{ii}}\,K profile of $-$320\,$\pm$\,10\,km\,s$^{-1}$ is consistent with that of the 2019 June Ca\,{\textsc{ii}} triplet profiles with an average of $-$307\,$\pm$\,8\,km\,s$^{-1}$. The candidate emission feature detected in the 5169\,\AA\ Fe\,{\textsc{ii}} region also changes in morphology between the two epochs, where the red-shifted component of the profile contracts to shorter wavelengths, corresponding to smaller velocities. We note that this region has also been shown to host 5167\,\AA, 5173\,\AA, and 5184\,\AA\,Mg\,{\textsc{i}} emission for some gaseous debris discs (see \citealt{melisetal20-1,dennihyetal20-2,fusilloetal20-1}). The potential O\,{\textsc{i}} emission feature appears to share the red-dominated asymmetry of the Ca\,{\textsc{ii}} triplet.

While the candidate 3934\,\AA\ Ca\,{\textsc{ii}}\,K and 8446\,\AA\ O\,{\textsc{i}} features discovered at HE\,1349--2305 appear to show morphologies consistent with the Ca\,{\textsc{ii}} triplet, this may not be the case for the 5169\,\AA\ Fe\,{\textsc{ii}} feature, and is not necessarily always true for emission lines formed by gaseous debris discs. The varied morphologies of different emission lines, both between atomic species and transitions of the same species, have been observed previously for other gaseous debris discs, most notably SDSS\,J1228+1040 \citep{manseretal16-1}. It was speculated for the disc around SDSS\,J1228+1040 that the distribution of emitting ions in the disc may differ for a given atomic transition. However, additional high signal-to-noise ratio spectroscopy is needed to confirm the presence of the additional emission features identified in Figs\,\ref{f-CaK+Fe_spec}\,\&\,\ref{f-OI} for the disc around HE\,1349--2305.

\section{Short-cadence observations of the Ca\,{\textsc{ii}} triplet}\label{s-shortterm}

\begin{figure}
\centerline{\includegraphics[width=0.5\textwidth]{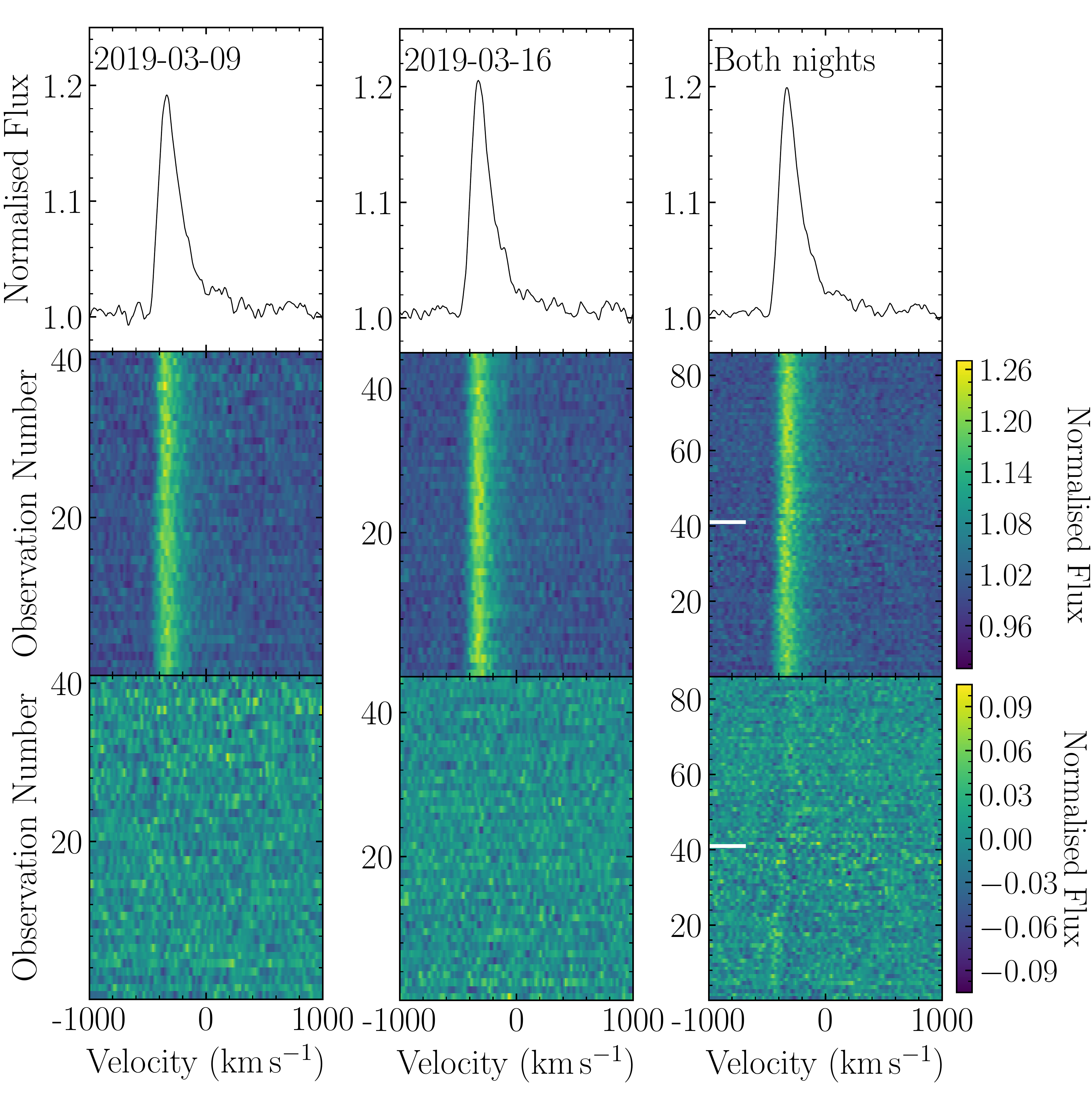}}
\caption{\label{f-spectrograms} Short-cadence (4\,min exposure time) GMOS spectra of the Ca\,{\textsc{ii}} triplet emission from the disc around HE\,1349--2305, where the triplet profiles have been co-added in velocity space to boost the signal-to-noise ratio. The figures show data from 2019 March 09, 2019 March 16, and both nights combined in the respectively labelled figures. We assume the rest air-wavelength for each transition. The top panels show the Ca\,{\textsc{ii}} triplet profiles averaged over the time-resolved data shown in the middle panels as trailed spectrograms, where the Ca\,{\textsc{ii}} triplet emission profiles can be seen as vertical streaks. The white bars show the change from data taken on 2019 March 09 to 2019 March 16. By subtracting the the average spectra (top panels) from the trailed spectrograms (middle panels), the bottom panels are obtained showing the average-subtracted trailed spectrograms. No short-term variability is detected, however in the average-subtracted trailed spectrogram including both nights of data there is a subtle signature of the precession ($\simeq$\,5\,$^{\textrm{o}}$) changing the morphology of the Ca\,{\textsc{ii}} triplet profiles between the two nights.}
\end{figure} 

\begin{figure*}
\centerline{\includegraphics[width=2\columnwidth]{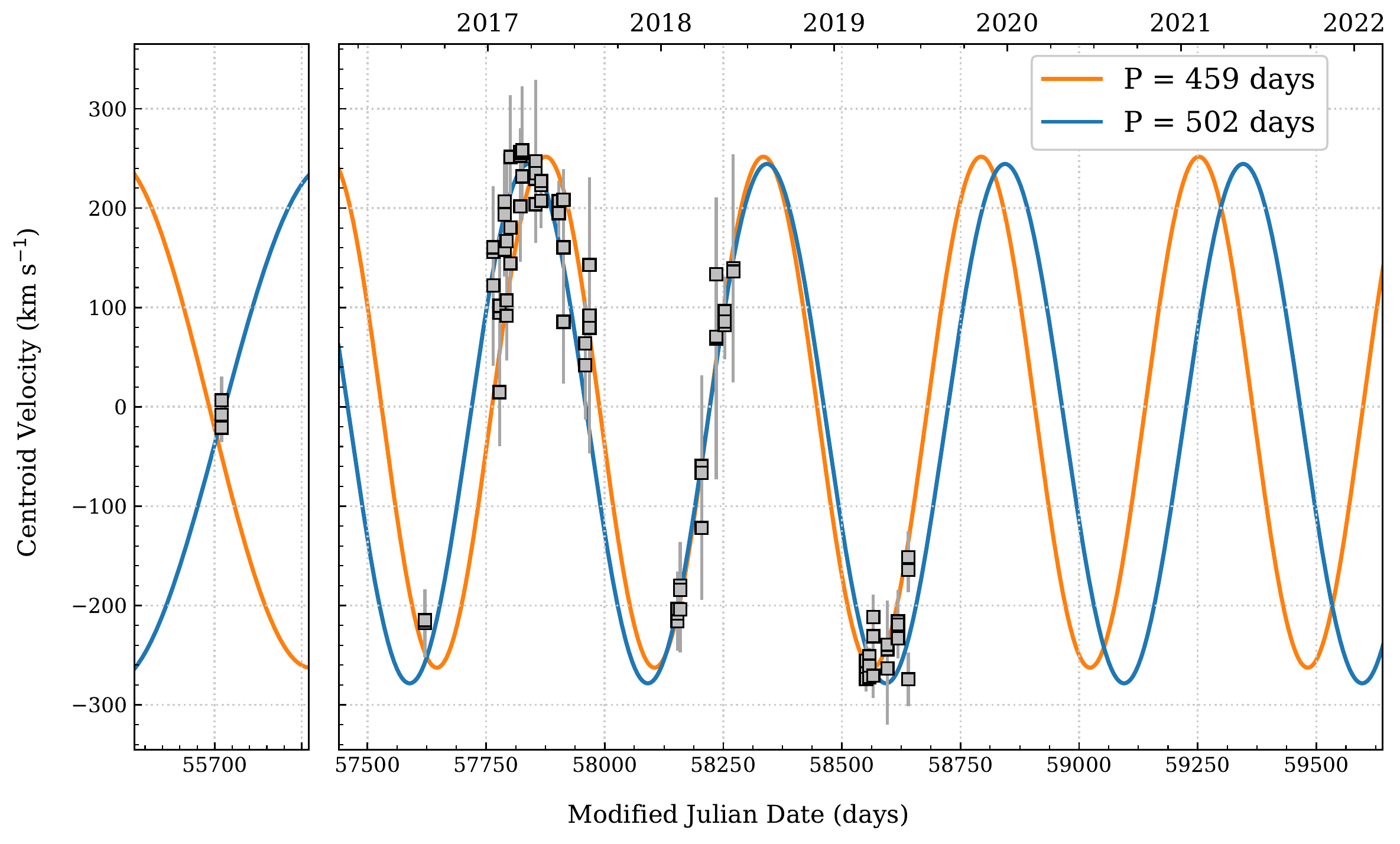}}
\caption{\label{f-cent1} Sinusoidal fits to the velocity centroids of the Ca\,{\textsc{ii}} triplet profiles shown in Fig.\,\ref{f-caii_profiles}, corresponding to periods of 459\,d (orange) and 502\,d (blue) obtained using the ANOVA method. Each observation (at a given time) is represented by three gray points (sometimes overlapping), corresponding to the three components of the Ca\,{\textsc{ii}} triplet. The $x$-axis is given in both Modified Julian date (bottom) and in calender year (top).}
\end{figure*}

We obtained a total of 86, 4\,min exposures (totalling 344\,min, or 5.73\,hr) on 2019 March 9 (41 exposures) and 16 (45 exposures) using the 8.1\,m Gemini South telescope to search for any short-term variations in the emission strength and morphology of the Ca\,{\textsc{ii}} triplet on the time-scale of the orbital period in the disc ($\simeq$\,hours). Periodic variations in the Ca\,{\textsc{ii}} triplet emission were observed at the debris disc around SDSS\,J1228+1040 and were found to be due to a planetesimal embedded in the disc with an orbital period of 123.4\,min \citep{manseretal19-1}. HE\,1349--2305 may host a disc-embedded planetesimal detectable through short-term periodic variations, however, the order of magnitude difference in the strength of the emission lines between SDSS\,J1228+1040 and HE\,1349--2305 makes this a challenging task. If the level of emission produced by the short-term signal is proportional to the brightness of the Ca\,{\textsc{ii}} triplet lines, then the signal strength of $\simeq$\,10\,per\,cent observed for SDSS\,J1228+1040 (e.g. fig.\,S4 of \citealt{manseretal19-1}) would correspond to an expected signal strength of $\lesssim$\,1\,per\,cent for the emission lines at HE\,1349--2305. The required signal-to-noise ratio to detect such a signal would be \textgreater\,100.

We co-added the normalised spectra of all three Ca\,{\textsc{ii}} triplet components in velocity to boost the signal-to-noise ratio, and present trailed spectrograms in Fig.\,\ref{f-spectrograms}. The typical signal-to-noise ratio of the velocity spectra is $\simeq$\,70, and the single night data show no evidence of any short term variations in the emission ruling out short-term variability at a level of 1.4\,per\,cent. If the expected short-term signal is proportional to the brightness of the Ca\,{\textsc{ii}} triplet then the quality of the data obtained is not sufficient to exclude the presence of a planetesimal similar to that around SDSS\,J1228+1040. We note that the trailed spectrogram combining both nights of data reveals very subtle variations in each of the Ca\,{\textsc{ii}} triplet profiles, which is a result of the apsidal precession of the debris disc between the two nights.

\section{Discussion}

We have presented updated monitoring of the long-term morphological variations seen in the emission profiles of HE\,1349--2305 and analysed the period and systemic velocity measurements obtainable from this data. We also searched for short-term variations in the strength and shape of the line profiles on the order of hours to no avail.

\subsection{Precession of the debris disc}

The relatively short precession period of HE\,1349--2305 makes this an ideal system for studying cycle-to-cycle stability. While SDSS\,J1228+1040 presents the strongest emission lines identified so far, it would take $\simeq$\,50\,yr to cover two precession cycles. 

While there appears to be some discrepancy in our fitting between early observations of HE\,1349--2305 in 2009 and 2011, and the later monitoring between 2016 and 2019, the morphological variations expected for a fixed intensity pattern are well described by the Doppler maps shown in Fig\,\ref{f-doppler_464}. Additional spectroscopy of the Ca\,{\textsc{ii}} triplet profiles is required to resolve the aliasing in the periods of variability we determine. Fig.\,\ref{f-cent1} shows sinusoidal curves with periods of 459\,d and 502\,d fit to the velocity centroid data, illustrating the drift between the two periods over time. By mid 2021 and beyond, the two periods will be separated enough that a single observation may be able to break the degeneracy, however this relies on the assumption that the apsidal precession has a fixed period. We therefore suggest that continued ($\simeq$ monthly) monitoring of this system over the next few years with high signal-to-noise ratio, moderate-resolution spectroscopy (resolving power R\,$\sim$\,10\,000) is required to break the period degeneracy and search for any robust deviation from a constant period over multiple precession cycles.

A useful property that can be obtained from the Doppler map is the distribution of emitting gaseous debris in velocity space, which can be compared against disc models and expectations from hypothesised disc configurations. The disc around SDSS\,J1228+1040 is thought to be moderately eccentric ($e$\,$\simeq$\,0.3-0.5), which is possibly generated by an embedded planetesimal. The eccentricity of the disc is likely driving the long-term precession through general relativistic precession and pressure forces \citep{miranda+rafikov18-1}. The emission profiles around HE\,1349--2305 appear significantly more asymmetric than those of SDSS\,J1228+1040, and along with the dramatically shorter precession period would indicate the disc is at least as eccentric as SDSS\,J1228+1040. For the 495\,d period Doppler map, we plotted an eccentric orbit in velocity space shown in Fig.\,\ref{f-eccentric} using the equations described in Appendix\,\ref{appendix-1}, which show that eccentric orbits are represented by circles in velocity space, offset from the origin with a displacement depending on $e$. The emission from the debris disc can be well described by an arc of an eccentric orbit, with the emission brighter at the apocentre of the orbit. We note that this is not a fit to the data, but a guide to the eye as a proof of concept, using an eccentricity, $e$\,=\,0.35, orbital period $P_{\textrm{orb}}$\,=\,60\,min, a disc inclination, $i$\,=\,1\,rad\,$\simeq$\,57.3\,$^{\textrm{o}}$, and $M_{\textrm{WD}}$\,=\,0.673\,\Msun\ \citep{koesteretal05-2,vossetal07-1}. The value of $i$ used here is the average value for the inclination of a randomly oriented disc given no information, and we use it here as an example. The pericentre value obtained for this set-up of $\simeq$\,0.24\,\Rsun\ is loosely consistent with the findings of \cite{miranda+rafikov18-1}. However, we refrain from drawing any definitive conclusions from this set-up given that multiple free parameters can be adjusted to obtain a similar fit (e.g. $e$\,=\,0.3, $P_{\textrm{orb}}$\,=\,120\,min, $i$\,=\,90\,$^{\textrm{o}}$).

\begin{figure}
\centerline{\includegraphics[width=1\columnwidth]{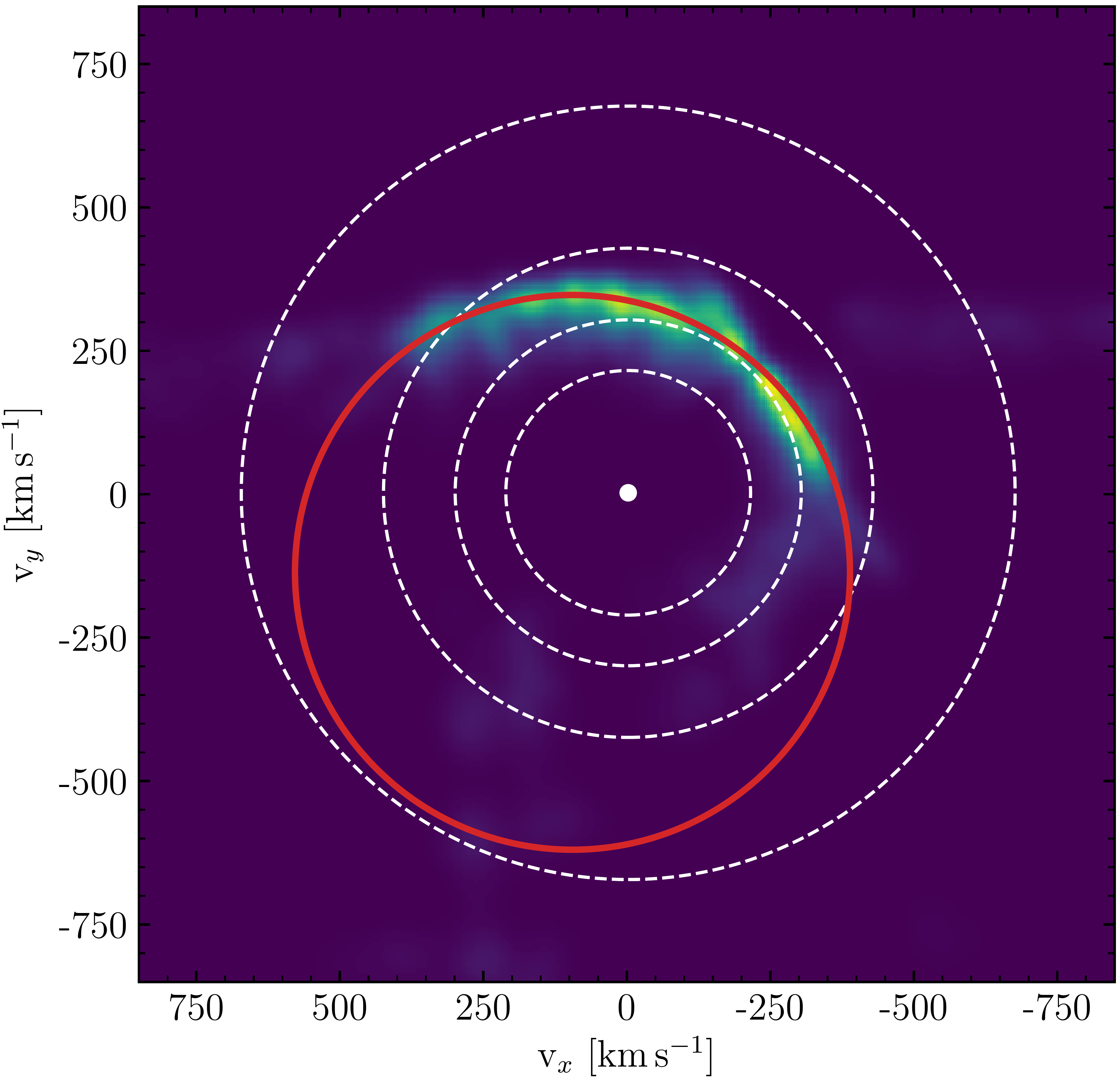}}
\caption{\label{f-eccentric} The 495\,d doppler map shown in Fig.\,\ref{f-doppler_464} with an eccentric orbit (red) overplotted with $e$\,=\,0.35, orbital period $P_{\textrm{orb}}$\,=\,60\,min, and an inclination, $i$\,=\,1\,rad\,$\simeq$\,57.3\,$^{\textrm{o}}$. We note that eccentric orbits are represented by offset circles in velocty space (see Appendix\,\ref{appendix-1}). The white dashed lines show circular orbits with radii 0.2, 0.5, 1.0, and 2.0 \Rsun\ in order from the largest circle to the smallest, assuming the same inclination.}
\end{figure} 

An interesting and simple consequence of gaseous material following an eccentric orbit is that the gas will be changing in radial distance from the white dwarf as it travels along the orbit. Photo-ionisation models show that the emission strength of a particular atomic species in a gaseous disc is dependent on the disc temperature (\citealt{melisetal10-1, kinnear11, gaensickeetal19-1}, see also discussion of \citealt{fusilloetal20-1}) - itself dependent on the distance from the white dwarf \citep{steeleetal20-1}. As this distance is varying throughout the eccentric orbit, we would expect the emission strength of the Ca\,{\textsc{ii}} triplet to change as well. For HE\,1349--2305, it appears that the material is brightest at apocentre where the material has the lowest velocities in the map. This is consistent with SDSS\,J1228+1040, where the brightest part of the Ca\,{\textsc{ii}} triplet intensity distribution in the disc occurs in the lowest velocity region - around apocentre (fig. 5 of \citealt{manseretal16-1}). This inverse correlation between emission strength and disc temperature has been observed in gaseous debris disc simulations, and is due to a shift in the ionisation balance of Ca\,{\textsc{ii}}/Ca\,{\textsc{iii}} \citep{hartmannetal11-1, hartmannetal16-1}, where photons of energy 11.87\,eV or higher are required for ionisation of Ca\,{\textsc{ii}} \citep{ferlandetal17-1}. 

The scenario described above also explains why the different atomic species observed in the disc at SDSS\,J1228+1040 have very different profile shapes. The morphology of the O\,{\textsc{i}} emission from the disc at SDSS\,J1228+1040 is almost in anti-phase with the shape of the Ca\,{\textsc{ii}} triplet, which can be explained by O\,{\textsc{i}} emission arising from hotter material closer to pericentre (see also fig.\,7.1 of \citealt{manser18-1}). This is consistent with the upper energy levels that produce the $\simeq$7775\,\AA\ triplet and 8446\,\AA\ O\,{\textsc{i}} lines ($\simeq$\,10.74\,eV and 10.99\,eV respectively), requiring a hotter gas orbiting closer to the white dwarf than compared with the Ca\,{\textsc{ii}} triplet (upper energy levels of $\simeq$\,3.12-3.15\,eV, \citealt{kramidaetal19-1}). However, the potential O\,{\textsc{i}} emission we detect for HE\,1349--2305 appears to share the morphology of the Ca\,{\textsc{ii}} triplet, suggesting they are being generated by gas in a similar location. A possible explanation for this is that given the potential for shorter orbital periods outlined above, the gaseous debris may be hotter around HE\,1349--2305, and both Ca\,{\textsc{ii}} and O\,{\textsc{i}} are more readily ionised (O\,{\textsc{i}} has an ionisation energy of 13.62\,eV) near the pericentre of the orbit and reducing the available ions for emission. However, this is largely speculation, and we have not taken into account the lower $T_{\textrm{eff}}$ of HE\,1349--2305 (18\,173\,K, \citealt{melisetal12-1}) compared with that of SDSS\,J1228+1040 (20\,713\,K \citealt{koesteretal14-1}). The production of Doppler maps (with a constant $\gamma$ and $P$) of multiple atomic species for a single gaseous debris disc would allow these postulates to be tested, as well as provide useful comparisons for models of the disc, such as those that have been produced using \texttt{CLOUDY} \citep{kinnear11, ferlandetal17-1, gaensickeetal19-1, steeleetal20-1}.

\subsection{Non-detection of short-term periodic variations}

We do not detect short-term variability due to the presence of a planetesimal, but cannot rule out variability below $\simeq$\,1.4\,per\,cent. While reducing the uncertainty of future observations of HE\,1349--2305 may help to rule out variability at a more significant level, it would require dedicated observations on an 8\,-\,10\,meter class telescope over multiple nights. If the emission generated by a single planetesimal is proportional in strength to the total emission from the gaseous component of the disc, then it would be more efficient to observe gaseous debris discs with emission features comparable in strength to those at SDSS\,J1228+1040. 

\cite{melisetal20-1}, \cite{dennihyetal20-2} and \cite{fusilloetal20-1} recently reported the combined detection of 14 new gaseous debris discs (for a total of 21 gaseous debris disc systems), five of which -- SDSS\,J0006+2858, WD\,0347+1624, WD\,0611--6931, WD\,J0846+5703, and WD\,J2133+2428, show strong Ca\,{\textsc{ii}} triplet emission. Furthermore, the morphology of the emission profiles of WD\,0347+1624 appears to vary on a time-scale of 2\,-\,3\,yrs; similar to HE\,1349--2305 and indicative of the precession of the debris disc. These systems are currently the best gaseous debris discs to probe in search of short-term variability on the time-scale of the orbital period in the disc induced by the presence of a single planetesimal.

While our discussion has focused on detecting the signal from a single planetesimal, it has been suggested that multiple planetesimals undergoing collisional cascades can explain the observed variability of the infrared components of white dwarf debris discs and their connection to gaseous emission \citep{swanetal21-1}. A debris disc hosting multiple planetesimals is not at odds with the detection of a single planetesimal at SDSS\,J1228+1040, as the majority of the mass in these discs are contained within the few largest bodies. Further research into the evolution of the gaseous component to white dwarf debris discs, the variability of the dusty infrared emission, and the survival of planetary bodies within the typically assumed Roche radius ($\simeq$\,1\,\Rsun) are required to refine these hypotheses.

\section{Conclusions}

We presented observations revealing continued morphological evolution of the emission from the gaseous debris disc around the white dwarf HE\,1349--2305. The Ca\,{\textsc{ii}} triplet profiles are well fit by a fixed, asymmetric intensity pattern undergoing apsidal precession. While we have determined aliased periods that require future observations to distinguish, we show that they are consistent with previous analysis of the velocity centroids of the profiles. We also hypothesise that the asymmetric intensity distributions seen in Doppler maps of HE\,1349--2305 and SDSS\,J1228+1040 are due to the radially dependant excitation conditions as the gas moves on eccentric orbits around the photo-ionising white dwarf. 

We find no evidence of short-term variability on the orbital time-scale of the disc, which may have indicated the presence of a planetesimal. However, if there is a correlation between the strength of the overall Ca\,{\textsc{ii}} triplet profile and any short-term signal, then our results are likely dominated by noise. Future observations of bright white dwarfs with strong emission from a gaseous debris disc, such as WD\,J0347+1624 or WD\,J0846+5703, would help to determine if this is the case. 

Gaseous debris discs around white dwarfs show a remarkable range of variability, and the potential to Doppler map these apsidally precessing discs using Doppler tomography is still in its early stages. Additional monitoring of the now 21 known gaseous debris discs is critical in following any morphological evolution present, and to eventually produce Doppler maps of them using Doppler tomography. These Doppler maps, along with the plethora of emission signatures beyond the Ca\,{\textsc{ii}} triplet seen in some systems will be vital in determining fundamental parameters of these discs, such as their temperature and surface density profiles, and studying their dynamical evolution.

\section*{Acknowledgements}

We thank the anonymous referee for their helpful comments that improved this manuscript. CJM, BTG, and TRM were supported by the UK STFC grant ST/T000406/1. The authors acknowledge financial support from Imperial College London through an Imperial College Research Fellowship grant awarded to CJM. BTG and TRM were supported by Leverhulme Research Fellowships. DV gratefully acknowledges the support of the STFC via an Ernest Rutherford Fellowship (grant ST/P003850/1). Based on observations obtained at the Southern Astrophysical Research (SOAR) telescope, which is a joint project of the Minist\'{e}rio da Ci\^{e}ncia, Tecnologia, Inova\c{c}\~{o}es e Comunica\c{c}\~{o}es (MCTIC) do Brasil, the U.S. National Optical Astronomy Observatory (NOAO), the University of North Carolina at Chapel Hill (UNC), and Michigan State University (MSU). Based on observations (Program ID: GTC1-16ITP) made with the Gran Telescopio Canarias (GTC), installed in the Spanish Observatorio del Roque de los Muchachos of the Instituto de Astrof\'isica de Canarias, in the island of La Palma. Data for this paper have been obtained under the International Time Programme of the CCI (International Scientific Committee of the Observatorios de Canarias of the IAC). Based on observations collected at the European Southern Observatory under ESO programmes 087.D-0858(A), 5100.C-0407(C), and 5100.C-0407(I). Based on observations obtained at the international Gemini Observatory under program number GS-2019A-Q-231. Gemini Observatory is a program of NSF’s OIR Lab, which is managed by the Association of Universities for Research in Astronomy (AURA) under a cooperative agreement with the National Science Foundation on behalf of the Gemini Observatory partnership: the National Science Foundation (United States), National Research Council (Canada), Agencia Nacional de Investigaci\'{o}n y Desarrollo (Chile), Ministerio de Ciencia, Tecnolog\'{i}a e Innovaci\'{o}n (Argentina), Minist\'{e}rio da Ci\^{e}ncia, Tecnologia, Inova\c{c}\~{o}es e Comunica\c{c}\~{o}es (Brazil), and Korea Astronomy and Space Science Institute (Republic of Korea). Also based on observations attained with the 6.5 m Magellan Telescopes located at Las Campanas Observatory, Chile.

\section*{Data availability}
All data underlying this article is publicly available from the relevant observatory archive (see program IDs in text) or will be shared on reasonable request to the corresponding author.

\appendix

\section{Eccentric orbits in velocity space}\label{appendix-1}

One curious property of eccentric orbits in velocity space is that they are represented as offset circles, which was identified independently by both \cite{mobius1843-1} and \cite{hamilton1847-1}, and has been discussed in numerous other works since (e.g., \citealt{carinenaetal16-1} and references therein), where velocity diagrams are usually referred to as `hodographs'. We re-illustrate here that eccentric orbits can be represented as offset circles in velocity space using the formalism described in Appendix\,A of \cite{manseretal16-2}. We start with their equations A12-15:

\begin{equation}
v_x  =  v_r \cos f - v_f \sin f, \label{eq:vx}
\end{equation}

\begin{equation}
v_y  =  v_r \sin f + v_f \cos f,\label{eq:vy}
\end{equation}

\begin{equation}
v_r  = \frac{e l \sin f}{\left(1 + e \cos f\right)^2} \,
\frac{\textrm{d}f}{\textrm{d}t},\label{eq:vr}
\end{equation}

\begin{equation}
v_f  = \frac{l}{1 + e \cos f} \, \frac{\textrm{d}f}{\textrm{d}t},\label{eq:vv}
\end{equation}

\noindent where $v$ is the velocity, $e$ is the eccentricity, $f$ is the true anomaly, $l = a(1-e^2)$ is the semi-latus rectum, $a$ is the semi-major axis, and $\frac{\textrm{d}f}{\textrm{d}t}$ is the derivative of the true anomaly with respect to time which can be written as

\begin{equation}
\frac{\textrm{d}f}{\textrm{d}t} = \frac{ (1+e\cos f)^2}{\left(1-e^2\right)^{3/2}} n. \label{eq:nud}
\end{equation}

\noindent where $n = 2\pi/P_{\textrm{orb}}$, and $P_{\textrm{orb}}$ is the orbital period. Substituting equation\,\ref{eq:nud} into equations\,\ref{eq:vr}\,\&\,\ref{eq:vv} results in

\begin{equation}
v_r  = \frac{e n l \sin f}{\left(1-e^2\right)^{3/2}},    \label{eq:vr2}
\end{equation}

\begin{equation}
v_f  = \frac{n l (1 + e \cos f)}{\left(1-e^2\right)^{3/2}}.\label{eq:vv2}
\end{equation}

\noindent Substituting equations\,\ref{eq:vr2}\,\&\,\ref{eq:vv2} into equations\,\ref{eq:vx}\,\&\,\ref{eq:vy} and reducing gives

\begin{equation}
v_x  =  - \frac{n l \sin f}{\left(1-e^2\right)^{3/2}}   , \label{eq:vx2}
\end{equation}

\noindent and

\begin{equation}
v_y  =  \frac{e n l}{\left(1-e^2\right)^{3/2}}  + \frac{n l \cos f}{\left(1-e^2\right)^{3/2}} .\label{eq:vy2}
\end{equation}

\noindent Finally, by substituting,

\begin{equation}
K = \frac{n l}{\left(1-e^2\right)^{3/2}} = \frac{2 \pi l}{P_{\textrm{orb}} \left(1-e^2\right)^{3/2}} = \frac{2 \pi a}{P_{\textrm{orb}} \left(1-e^2\right)^{1/2}} , \label{eq:K}
\end{equation}

\noindent into equations\,\ref{eq:vx2}\,\&\,\ref{eq:vy2}, we get

\begin{equation}
v_x  = - K \sin f   , \label{eq:vx3}
\end{equation}

\noindent and

\begin{equation}
v_y  = K e  + K \cos f .\label{eq:vy3}
\end{equation}

\noindent Equations\,\ref{eq:vx3}\,\&\,\ref{eq:vy3} show that an eccentric orbit in velocity space is represented by a circle of radius $K$ and offset from the origin by $Ke$. $K$ can be modified easily to include the inclination, $i$ (given that $v_{\textrm{obs}} = v \sin i$), as,

\begin{equation}
K = \frac{2 \pi a \sin i}{P_{\textrm{orb}} \left(1-e^2\right)^{1/2}} , \label{eq:K2}
\end{equation}

\noindent and can be rewritten in terms of $P_{\textrm{orb}}$, $M_{\textrm{WD}}$, $e$, and $i$,

\begin{equation}
K = \left(\frac{2 \pi G M_{\textrm{WD}}}{P_{\textrm{orb}}}\right)^{1/3} \frac{\sin i}{\left(1-e^2\right)^{1/2}} , \label{eq:K3}
\end{equation}

\noindent where $G$ is the gravitational constant.



\bibliographystyle{mnras}
\bibliography{aamnem99,aabib}


\bsp	
\label{lastpage}
\end{document}